\documentstyle[epsf,amsfonts,aps]{revtex}
\begin{document}
\title{One-Dimensional Electron Liquid in an Antiferromagnetic Environment:\\
Spin Gap from Magnetic Correlations}

\author{Mats Granath and Henrik Johannesson}

\address{Institute of Theoretical Physics, Chalmers University of 
Technology and 
G\"oteborg University, SE 412 96 G\"oteborg, Sweden}

\maketitle
\begin{abstract}
We study a one-dimensional electron liquid coupled by a weak 
spin-exchange interaction to an antiferromagnetic spin-$S$ ladder with 
$n$ legs. A 
perturbative renormalization group analysis in the semiclassical limit 
reveals the opening of a spin gap, driven by the local magnetic 
correlations on the ladder. The effect, which we argue is present for 
{\em any} gapful ladder {\em or} gapless ladder with $nS\gg 1$, is enhanced 
by the repulsive interaction among the conduction electrons but is insensitive 
to the sign of the spin exchange interaction with the ladder. Possible 
implications for the striped phases of the cuprates are discussed. 
\end{abstract}


\newcommand{\vJ}{\mbox{\boldmath $J$}}             
\newcommand{\vS}{\mbox{\boldmath $S$}}             
\newcommand{\vl}{\mbox{\boldmath $\ell$}}
\newcommand{\vn}{\mbox{\boldmath $n$}}
\newcommand{\vsigma}{\mbox{\boldmath $\sigma$}}
\newcommand{\del}{\partial}
\newcommand{\LR}{{}^L_R}                           
\newcommand{\no}[1]{: \! #1 \! :}                  
\newcommand{\Z}{{\Bbb Z}}                          

The coexistence of conducting electrons and localized spins remains
one of the 
most challenging problems of condensed matter physics, as evidenced
by the 
enormous effort put into the study of, say, the Kondo lattice 
or doped antiferromagnets \cite{rev}. The
recently discovered 
striped phases in La$_{2-x}$Sr$_{x}$CuO$_4$ 
and various other high-$T_c$ cuprates \cite{tranquada} add a new twist to
this class of problems. 
``Stripes'' is the name for spontaneously formed domain walls
across which the 
two-dimensional antiferromagnetic order in these materials changes
sign, and along which the doped holes 
are concentrated. The stripes
are slowly fluctuating 
structures and may locally be modeled as metallic wires - in fact,
Luttinger liquids \cite{voit} - embedded 
in an antiferromagnetic environment. 
As suggested by Emery, Kivelson and Zachar \cite{ekz},
pair-tunneling 
of holes between the stripes and the environment 
may produce an electronic spin gap favoring 
either a charge density wave or superconducting correlations.
Josephson coupling between stripes is 
expected to suppress the charge density wave, paving the way for 
superconductivity. Suggestions have also been made that a spin gap in
the striped phase may be identified with the ``normal-state'' pseudogap 
observed in the underdoped cuprates \cite{Tsuei}.

In this Letter we also consider a one-dimensional electron liquid in an
antiferromagnetic Mott insulating environment, and here
focus on the role of the
spin-exchange interaction between itinerant and localized
electrons. This problem belongs to the more general class of {\em Luttinger liquids
in
active environments} \cite{castroneto}, a topic of importance not only 
to the striped phases, but also to e.g. nanotube \cite{balents} and
Kondo chain physics \cite{sikkema}. 
It is important to realize that in the present case spin and momentum
conservation severely restrict the possible relevant interactions
between the electron liquid and its environment. In particular, since
the Fermi momentum of the Luttinger liquid (away from half-filling) is
incommensurate with that of any low-energy excitation of the Mott insulator
we can neglect as irrelevant terms which transfer single holes to the
insulator. Pair hopping is still allowed and is favored
when the spins in the
environment have a tendency to form singlets, as may be the case when there is a
large pre-existing spin gap in the environment \cite{ekz}. In addition, 
however, a spin
exchange interaction is always present, and is expected to become dominant 
for smaller gaps,
correlating with a smaller density of local spin singlets. This is the case we
consider here.

Treating the localized spins semiclassically, we
exploit a path integral formalism to construct a low-energy
effective action with a companion
set of perturbative renormalization group (RG) equations. Their
solution reveals the opening 
of an electronic spin gap on the stripes, driven by the magnetic
correlations in the 
environment. Rather strikingly, the effect is enhanced by the
repulsive electron-electron 
interaction, but is insensitive to whether 
the coupling to the
environment is ferro- or antiferromagnetic.
Although our approach allows for a fully 
controlled calculation only for large values of the localized spins
or - as we shall see - for sufficiently wide 
antiferromagnetic domains between the stripes, we shall argue that
our results are robust in the limit of narrow 
spin-1/2 domains, at least in the case when the environment is 
noncritical.   

As lattice model we take a Hubbard chain (representing a stripe)
coupled 
to the first leg of a neighboring spin ladder (representing the 
environment) by a spin-exchange interaction: 
\begin{eqnarray}
H&=&\sum_{r=1}^{N}[-t(c_{r+1,\sigma}^{\dagger}c_{r,\sigma} + h.c.) +
Un_{r,\sigma}n_{r,-\sigma}+
J_K c_{r,\sigma}^{\dagger}\vsigma_{\sigma\mu}
c_{r, \mu} 
\cdot \vS_{r,1} + \nonumber \\
&& J_H (\sum_{j=1}^{n_{leg}}\vS_{r,j} \cdot
\vS_{r+1,j} + 
\vS_{r,j} \cdot \vS_{r,j+1})] \, , \ \ \ \ \  \vS_{N+1,j}=\vS_{r,n_{leg}+1}=0, \ \
\ \ \ J_H>0\,. \label{Hamiltonian}
\end{eqnarray}
Here $c_{r,\sigma}$ is a conduction electron operator at site $r$
with spin index 
$\sigma$,    
$n_{r,\sigma}=c^\dagger_{r,\sigma} c_{r,\sigma}$ is a number 
operator, and
$\vS_{r,j}$  is the operator for the 
localized spin at the site with coordinates $r$ along legs and $j$
along rungs. 
The model can be extended to 
include a coupling to antiphase ladders on either side of the stripe. As long as
these ladders are correlated and the stripe is away from half-filling this 
will only change 
the magnitude of the couplings in the effective theory - to be derived 
below - but will not 
result in any qualitative changes \cite{mghj}. 

The model in (\ref{Hamiltonian}) can be taken as an {\em effective} model of a local stripe
phase in the cuprates, valid on length- and time scales set by the fluctuation dynamics of
the stripes (which is expected to be much slower than the dynamics of
charge carriers along the stripe). For the purpose of exploring whether a spin gap
opens up or not, we can count on the stripe as being metallic, as assumed in
(\ref{Hamiltonian}), since for weak disorder (induced e.g. by the dopant potentials) 
localization effects set in at length scales much larger than any relevant spin gap length
scale. We should point out that since our 
model has the presence of
stripes already built into it, the model cannot describe the instability that
triggers the striped phases. For this, one must turn to other approaches, as in
\cite{Low}.

Given the Hamiltonian in (\ref{Hamiltonian}), its partition function 
can be expressed as a Euclidean path integral by using 
coherent spin states in the semiclassical (large-$S$) limit of the 
localized spins, i.e. taking 
$\vS_{r,j} \longrightarrow S{\mbox{\boldmath 
$\Omega$}}_{r,j}$, where ${\mbox{\boldmath $\Omega$}}$ is a vector
of unit 
length. This gives
\begin{equation}
Z=\int {\cal D}[{\bf \Omega}]{\cal D}[c]{\cal 
D}[c^{\dagger}]
e^{-{\cal S}[{\bf \scriptstyle \Omega},c,c^\dagger]}\,, 
\label{Partitionfunction}
\end{equation}
with action
\begin{equation}
{\cal S} = \int d\tau \ iS \ \sum_{r,j}
\Phi_{Berry}[{\mbox{\boldmath $\Omega$}}_{r,j}(\tau)] 
+ \sum_{r} c^\dagger_{r,\sigma} \partial_{\tau} c_{r,\sigma} + 
H(c^\dagger, c,
S{\mbox{\boldmath $\Omega$}})\, . \label{Action}
\end{equation}
The first term in (\ref{Action}) is a sum over Berry phases, coming 
from the overlap of the coherent spin states:
$\Phi_{Berry}[{\mbox{\boldmath $\Omega$}}_{r,j}(\tau)] = \int_0^1 \
du \
{\mbox{\boldmath $\Omega$}}_{r,j}(u,\tau)
\cdot [\partial_u {\mbox{\boldmath $\Omega$}}_{r,j}(u,\tau) \times
\partial_{\tau}
{\mbox{\boldmath $\Omega$}}_{r,j}(u,\tau)]\,,$
where ${\mbox{\boldmath $\Omega$}}_{r,j}(u\!=\!1,\tau) = 
{\mbox{\boldmath $\Omega$}}_{r,j}(\tau)$, ${\mbox{\boldmath 
$\Omega$}}_{r,j}(u\!=\!0,\tau) = \mbox{constant}$, with $u$ a dummy
variable. The Hamiltonian term $H(c^\dagger, c, 
S{\mbox{\boldmath $\Omega$}})$ in (\ref{Action}) acts at time slice $\tau$
and is obtained from (\ref{Hamiltonian}) by substituting electron and 
spin operators by corresponding Grassmann fields 
$(c^{\dagger}, c)$ and 
classical vectors $S{\mbox{\boldmath $\Omega$}}$, respectively.
For the purpose of formulating a low-energy theory we linearize the
electron spectrum close to the Fermi points $\pm k_F$,
assuming  that
$U\ll \epsilon_F=2t(1-\cos ak_F)$, and set $c_{r,\sigma} = \sqrt{a/2\pi}
(e^{-ik_Far} \psi_{L\sigma}(ar) +
e^{ik_Far}
\psi_{R\sigma}(ar)),$
with $a$ the lattice spacing, $n_e$ the electron density,
$k_F=n_e\pi/2a$, and $\psi_{L/R\sigma}$ left/right moving chiral fields.

We expect that short-range antiferromagnetic correlations are
present on 
the ladder also at the quantum level, implying that the partition 
function at low energies is dominated by paths with 
\begin{equation}
{\mbox{\boldmath $\Omega$}}_{r,j} = 
[(-1)^{r+j}\sqrt{1-{\mbox{\boldmath 
$\ell$}}_{r,j}^2/S^2}\, \vn_r 
+ \vl_{r,j}/S)]\,, \label{Haldane}
\end{equation}
where $\vn_{r}\cdot\vl_{r,j}=0$ and $|\vn|=|{\mbox{\boldmath 
$\Omega$}}|=1$. Here $\vn$ is the local N\'eel-order 
parameter field, while 
$\vl/S$ represents small fluctuations
of the local 
magnetization \cite{haldane}. For this to be a viable description of the ladder we
require 
that the coupling to the conduction electrons is small, i.e. $|J_K|\ll 
J_H$, and also assume that the
antiferromagnetic correlation length along the legs is much greater
than the width of the 
ladder, allowing for $\vn$ to be taken constant along the rungs 
\cite{dellAringa}.

We first consider the case of free electrons $(U=0)$, away from half-filling  
($n_e \neq 1$). 
Taking the continuum limit of (\ref{Action}) and neglecting terms of higher 
than 
quadratic order in $\vl$ and $\partial_{\mu}\vn$,
one obtains the action
\begin{eqnarray}
{\cal S}&=& \int dx d\tau \left[ 2\pi 
iS\sum_{j}(-1)^j(\frac{1}{4\pi}\vn \cdot 
(\partial_{\tau} \vn \times \partial_{x} 
\vn)) - \frac{i}{a} (\vn
\times \partial_{\tau} \vn) \cdot \sum_{j}
\vl_j \nonumber  \right.\\
& + & \frac{1}{2\pi} \bar{\psi}(\gamma^0 \partial_{\tau} + \gamma^1
v_F \partial_x)\psi + \frac{J_K}{\pi} (\vJ_L+
\vJ_R ) \cdot \vl_1 \nonumber \\
&+& \left.
\frac{aJ_H}{2} n_{leg} S^2 (\partial_x \vn)^2 
 +  \frac{J_H}{a} \sum_{j}^{} (\frac{5}{2}\vl_j^2+ 
\frac{1}{2} \vl_{j+1}^2 + \vl_j \cdot \vl_{j+1})\right]\,, 
\label{lowenergyaction}
\end{eqnarray}
with spin currents 
$\vJ_{\LR}=\,\no{\frac{1}{2}\psi^{\dagger}_{\LR\sigma}\vsigma_{\sigma\mu}
\psi_{\LR\mu}}$, $\psi=(\psi_L,\psi_R)^T$ a Dirac fermion with velocity
$v_F=2at\sin ak_F$, and with $\gamma^0=\sigma^x$, $\gamma^1=\sigma^y$, 
$\bar{\psi}=\psi^{\dagger}\gamma^0$. 
The Gaussian integral over $\vl$ in the partition function of 
(\ref{lowenergyaction})
can be carried out by means of the substitution
$\vl'_i=\vl_i+L^{-1}_{ij}{\mbox{\boldmath $\omega$}}_j$, with 
$L_{ij}=J_H\delta_{ij}(6-\delta_{i1}-\delta_{in_{leg}})+J_H\delta_{ij\pm 1}$
and ${\mbox{\boldmath $\omega$}}_j=-i(\vn\times\partial_{\tau}\vn)+
\delta_{j1}\frac{J_Ka}{\pi}\vJ_{\perp}$, where we define 
$\vJ_{\perp}\equiv\vJ-(\vJ\cdot\vn)\vn$ with $\vJ\equiv\vJ_L+\vJ_R$.
We have here used the identity 
$\vJ\cdot\vl=(\vJ-(\vJ\cdot\vn)\vn)\cdot\vl$ 
to preserve the constraint 
$\vn(x)\perp \vl_j(x)$ in the substitution $\vl\rightarrow \vl'$, 
an observation crucial to the subsequent analysis of the problem.

This gives
\begin{eqnarray}
{\cal S} = {\cal S}_{NL\sigma} + {\cal S}_{Dirac} + {\cal
S}_{I}\,, 
\end{eqnarray}
where
\begin{eqnarray}
{\cal S}_{NL\sigma}&=& \frac{1}{2g}\int dx 
d\tau\left(\frac{1}{c}(\partial_{\tau}\vn)^2
+c(\partial_x\vn)^2\right)
+ 2\pi iS\sum_j(-1)^j \frac{1}{4\pi}\int dx d\tau \,\vn\cdot
(\partial_{\tau}\vn\times\partial_x\vn)\,,\label{sigma}\\ 
{\cal S}_{Dirac}&=& \frac{1}{2\pi}\int dx d\tau
        \bar{\psi}(\gamma^0 \partial_{\tau} + \gamma^1v_F
\partial_x)\psi
\label{Dirac}\,,\\
{\cal S}_{I}&=&\frac{1}{2\pi}\int dx d\tau\left(
        J_K C_1i(\vn\times\partial_{\tau}\vn)
                \cdot\vJ_{\perp} -
        \frac{aJ^2_K}{\pi}C_2\vJ_{\perp}\cdot\vJ_{\perp}
                \right)\,. \label{interaction} 
\end{eqnarray} 
Here ${\cal S}_{NL\sigma}$ is a nonlinear $\sigma$ model describing the ladder, 
with coupling $g^{-1}=S(J_Hn_{leg}\sum_{ij}L_{ij}^{-1})^{1/2}$ and velocity 
$c=aS(J_Hn_{leg}/(\sum_{ij}L^{-1}_{ij}))^{1/2}$,  
and with the topological 
term $2\pi iS\sum_j(-1)^j \frac{1}{4\pi}\int dx d\tau \vn\cdot
(\partial_{\tau}\vn\times\partial_x\vn)= i\theta Q$, where
$\theta\equiv 2\pi S\sum_{j=1}^{n_{leg}}(-1)^j$ is the topological angle and
$Q\,\epsilon\,\Z$ the winding number of the mapping $\vn: S^2 \rightarrow S^2$.
Note that the topological term is absent for even-leg ladders and also 
effectively for odd-leg ladders with integer spin, while for odd-leg 
ladders with half-odd-integer spin it is present with $\theta$ effectively
equal to $\pi$ \cite{khveshchenko}. We shall return to the implications of this below.
The Dirac action ${\cal S}_{Dirac}$ in (\ref{Dirac}) represents the 
electrons on the stripe, coupled to the ladder by 
${\cal S}_{I}$ in (\ref{interaction}), with $C_1=\sum_{i}L^{-1}_{i1}$ and 
$C_2=L^{-1}_{11}$.    

What is the effect of the interaction $S_{I}$? In particular, we wish to 
explore whether it may open up a spin gap for the electrons on the stripe.
For this 
purpose we shall treat the interaction $S_{I}$ by means of a perturbative RG 
approach, using a mean-field formulation of the local N\'eel-order parameter 
field $\vn$. Specifically, we will derive an
effective action for the spin sector which is valid over distances
over which the spin ladder is ordered. 
Within the limits of  
validity of this action we then integrate out the short wavelength
degrees of freedom to obtain its RG flow, allowing us to address the 
question above. 

Thus, given a patch in Euclidean space-time supporting local N\'eel order, we 
take the $\vn$-field to be in
a fixed (but arbitrary) direction $\tilde{\vn}$.
Introducing a local coordinate system ($x,y,z$) with $\hat{z}$ in the direction of
$\tilde{\vn}$,
and using the operator identity
$J^z_{\LR}J^z_{\LR}= \frac{1}{3}\vJ_{\LR}\cdot\vJ_{\LR}$,  
we obtain from (\ref{interaction}) $-$ dropping
the rapidly fluctuating first term of ${\cal S}_I$ \cite{footnote} $-$
an effective 
interaction $\tilde{{\cal S}}_{I}$ valid up to length-scales of the size of 
the ordered region,
\begin{equation}
\tilde{{\cal S}}_{I}=-\frac{g_J}{2\pi}\int dx d\tau
(\frac{2}{3}(\vJ_L\cdot\vJ_L+\vJ_R\cdot\vJ_R)+2(J_L^xJ_R^x+J_L^yJ_R^y))\,,
\label{int}
\end{equation}
with coupling $g_J=aJ_K^2C_2/\pi \approx aJ_K^2/4\pi J_H$. Note that the spin 
anisotropy of the induced interaction in 
(\ref{int}) is a direct consequence of the local N\'eel order of the $\vn$-field. 
Also note that the coupling $g_{\scriptscriptstyle J}$ is quadratic in $J_K$ and hence the same
for ferro- and antiferromagnetic spin exchange between the stripe and the
environment.

Bosonizing the Dirac action (\ref{Dirac}), i.e. splitting it into a 
charge boson and a (level $k=1$) 
Wess-Zumino-Witten ${\cal S}_{WZW,k=1}$ model for the spin degrees of 
freedom, we 
absorb the quadratic terms of (\ref{int}) into ${\cal S}_{WZW,k=1}$ 
via a Sugawara construction, thus obtaining an  
effective action $\tilde{{\cal S}}_{spin}$ for the spin sector of the conduction electrons:
\begin{equation}
\tilde{{\cal S}}_{spin}={\cal S}_{WZW,k=1}+\lambda_0 
\int d^2x(J_L^xJ_R^x+J_L^yJ_R^y)\,, \label{Seff}
\end{equation}
where $x^0=v_s\tau$, $v_s=v_F-2g_{\scriptscriptstyle J}$, and with dimensionless coupling
$\lambda_0=-g_{\scriptscriptstyle J}/\pi v_s$. Since $|\lambda_0|\ll 1$ we can
use standard perturbative RG techniques to analyze $\tilde{{\cal 
S}}_{spin}$, and at one-loop level we arrive at  
the scaling equations
\begin{equation}
\frac{d\lambda^i}{d\ln L}=2\pi\lambda^j\lambda^k\,\ \ \,,k\neq j\neq
i\,,
\label{scaling}
\end{equation}
for the couplings $\lambda^i$ of the operators $J_L^iJ_R^i$, with $L$ a 
short-distance cut-off. Using (\ref{scaling}) to solve for the RG flow, we 
obtain the trajectories $\lambda^2-(\lambda^z)^2=\lambda_0^2$ with 
$\lambda\equiv \lambda^x=\lambda^y$, and thus the scaling equation for 
$\lambda$: $d\lambda/d\ln L = 2\pi\lambda(\lambda^2-\lambda_0^2)^{1/2}$,
which upon integration gives $\arctan(\sqrt{(\lambda/\lambda_0)^2 -1})=
2\pi|\lambda_0|\ln L/a$.
Hence, $|\lambda|$ grows under renormalization
and at the length scale where $|\lambda|\sim{\cal O}(1)$ the
perturbative treatment breaks down. This scale - where the perturbation 
is of the same order 
of magnitude as the fixed point action and renders the theory 
non-critical - defines the correlation length $\xi_s$ of the electron 
spin sector. 
Using $|\lambda(\xi_s)|\sim{\cal O}(1)\gg|\lambda_0|$ in the scaling equation for $\lambda$
we thus obtain $\xi_s\approx ae^{1/4|\lambda_0|}$, with an associated spin gap
\begin{equation}
\Delta \approx\frac{v_s}{a}e^{-1/4|\lambda_0|}. \label{gap}
\end{equation}
The formation of a gap in this model is confirmed by 
the fact that (\ref{Seff}) corresponds to a fermionic low-energy 
formulation of a spin-$\frac{1}{2}$ XXZ chain (with a U(1)$\times$Z$_2$
symmetry)
\cite{AffleckLH}. The growing coupling constant scenario  
corresponds to an Ising anisotropy $J_z>1$ of the XXZ chain, for which the 
latter is known to have a N\'eel ordered groundstate with a broken Z$_2$ symmetry and
a mass gap.  

The procedure leading up to (\ref{gap}) 
requires that the environment exhibits N\'eel order over length scales 
exceeding $\xi_s$. Here we have to distinguish between spin ladders described by 
(\ref{sigma}) with vanishing topological term (even-leg and odd-leg 
ladders with integer spin) and those where the topological term is 
present with $\theta=\pi$ (odd-leg ladders with half-odd-integer spin).  
The behavior of the nonlinear $\sigma$ model without topological
term is well
established \cite{AffleckLH}; it has a finite mass gap and is ordered over 
distances given by the corresponding correlation length 
$\xi_{\sigma}$. In contrast, the behavior when $\theta=\pi$ is not
rigorously known, although the consensus is that the topological term 
drives a crossover to the critical k=1 WZW model at a length 
scale also set by $\xi_{\sigma}$ \cite{affleckhaldane}. 
However, in the weak coupling regime the topological term is effectively 
inactive \cite{chakravarty}, and as a consequence there is no distinction between  
gapless and gapful ladders on length-scales 
shorter than $\xi_{\sigma}$. 
It follows that the condition $\xi_s<\xi_{\sigma}$ validating our analysis  
is the same for gapless and gapful ladders. Evaluating $g$ we find
$g^{-1} \approx 0.36Sn_{leg}$, which in the weak-coupling regime  
with $\xi_{\sigma}\sim age^{2\pi/g}$ \cite{Sierra}
implies the consistency condition
\begin{equation}
0.3Sn_{leg} > \frac{J_Ht}{J_K^2} \gg 1 \,. \label{condition}
\end{equation}

While (\ref{condition}) shows that our perturbative RG calculation is  
well-controlled only for large spins or wide ladders, it is important to 
emphasize that the interaction $S_I$ in (\ref{interaction}) is well-defined 
for {\em any} values of $S$ or $n_{leg}$. As the symmetry of $S_I$ does 
not change when tuning the values of $S$ or $n_{leg}$, we expect that the 
result for the spin gap in (\ref{gap}) is analytic in these parameters 
with corrections that remain subleading as long as  
no topological effects intervene. On the other hand, 
when $\theta =\pi$, a 
violation of (\ref{condition}) may change the physics, as suggested by 
bosonization and DMRG results for the Heisenberg-Kondo lattice model 
($n_{leg} =1, S=1/2$) \cite{fujimoto,sikkema}: No gap is found for ferromagnetic coupling 
\cite{ners}
while for antiferromagnetic coupling the combined gap for itinerant and 
localized electrons scales as $\sim \mbox{exp}(-const.\,(\pi J_H/2 + 
v_F/J_K))$.   
It might be appropriate to add a note concerning the prospect that  
non-perturbative effects at length scales {\em larger} than 
$\xi_{\sigma}$ could possibly 
carry over to the electron liquid. Although we cannot rigorously exclude 
it, it seems improbable considering the fact that 
the spin sector of the electron liquid develops a mass at a length scale which
is shorter than and independent of 
$\xi_{\sigma}$ and as such the mass is already well established at the 
scale where non-perturbative effects from the ladder may come into play.  
  
Let us now include the electron-electron interaction in 
(\ref{Hamiltonian}) ($U\neq 0$).
At the level of the effective action for the electron spin sector
this changes $\tilde{{\cal S}}_{spin}$ in (\ref{Seff}) into
\begin{equation}
\tilde{{\cal S}}_{spin}={\cal S}_{WZW,k=1}+ 
\int d^2x\,\lambda_0(J_L^xJ_R^x+J_L^yJ_R^y)+\lambda^z_0J_L^zJ_R^z\,,
\label{}
\end{equation}
with renormalized velocity $v_s=v_F-2g_{\scriptscriptstyle J}-g_{\scriptscriptstyle U}$ and  
couplings
$\lambda_0=-(g_{\scriptscriptstyle J}+g_{\scriptscriptstyle U})/\pi v_s$, 
$\lambda^z_0=-g_{\scriptscriptstyle U}/\pi v_s$, where $g_{\scriptscriptstyle U} = aU/\pi$. 
Carrying out the RG analysis as above we obtain the spin gap
\begin{equation}
\Delta 
=\frac{v_s}{a}\exp(-\frac{\pi/2-\arctan(\lambda_0^z/\delta\lambda)}
{2\pi\delta\lambda})\,,
\end{equation}
where $\delta\lambda=\sqrt{\lambda_0^2-(\lambda_0^z)^2}$. Thus, as shown 
in  Fig.~1, a repulsive electron-electron interaction $(U>0)$ produces a 
larger gap, while for $U<0$ the outcome depends on the precise ratio 
between $g_{\scriptscriptstyle U}$ and $g_{\scriptscriptstyle J}$.
An interpretation of the surprising scenario of a decrease of the gap for $U<0$
due to the environment is that the competition between the attractive 
electron-electron interaction, which enhances on-site singlet pairing,
and the Ising anisotropy (discussed above), which enhances 
local N\'eel order, frustrates the system and hence reduces the
gap. It should, however, 
be noted that the 
actual vanishing of the gap at 
$g_{\scriptscriptstyle U}/g_{\scriptscriptstyle J}=-1$ cannot be rigorously 
concluded from our model as the self-consistency condition
$\xi_s<\xi_{\sigma}$ in this case requires 
$Sn_{leg} \rightarrow \infty$.


\begin{figure}
\centerline{\epsfbox{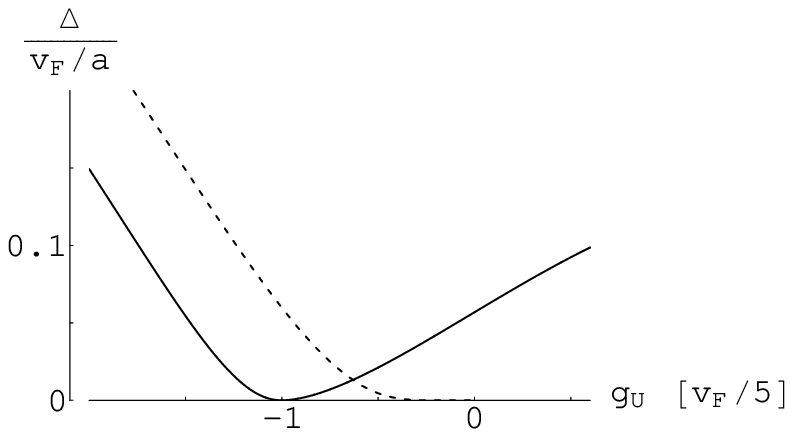}}
\caption{The spin gap $\Delta$ as a function of $g_{\scriptscriptstyle U}$;
 the solid line is for $g_{\scriptscriptstyle J}=v_F/5$ and
the dashed line for $g_{\scriptscriptstyle J}=0$.}
\end{figure}

In summary, we have shown that a one-dimensional electron liquid weakly coupled by a 
spin-exchange interaction to a spin ladder with $Sn_{leg} \gg 1$ develops a 
spin gap. The gap exhibits a strong dependence on the sign and magnitude 
of the itinerant electron-electron interaction, but is insensitive to whether 
the coupling to the ladder is ferro- or antiferromagnetic. A symmetry 
argument implies that these results hold for {\em any} gapful ladder {\em or}
gapless ladder with $Sn_{leg}\gg 1$.  
Applied to the striped phases seen in the cuprates this
may suggest that the local antiferromagnetic correlations in the insulating 
domains may conspire with
the electron correlations on the stripes to produce a sizeable spin gap. 
Details and extensions will be published elsewhere.

We wish to thank I. Affleck, S. A. Kivelson, A. A. Nersesyan, A. M. Tsvelik,
and J. Voit for
discussions and correspondence. H. J. acknowledges support from the Swedish 
Natural Science Research Council.


\begin{references}
\bibitem{footnote} In a Hamiltonian
formulation of the nonlinear $\sigma$ model \cite{Sierra}
$\vn\!\times\!\partial_t\vn$ is the angular momentum
density which is equivalent to the ferromagnetic component $\vl$ in 
Eq. (\ref{Haldane}). Since $\vl$ is rapidly varying
compared to $\vn$ it averages to zero over the length scales
considered here.
\end{references}
\end{document}